# "What if she doesn't feel the same?" What Happens When We Ask AI for Relationship Advice

**Niva Manchanda**[1] nmanchanda@ku.edu
**Akshata Kishore Moharir**[2] akshatankishore5@gmail.com
**Ratna Kandala**[1] ratnanirupama@gmail.com

[1]Department of Psychology, University of Kansas, USA
[2]Independent Researcher, Oregon, USA

## Abstract

Large Language Models (LLMs) are increasingly being used to provide support and advice in personal domains such as romantic relationships, yet little is known about user perceptions of this type of advice. This study investigated how people evaluate LLM-generated romantic relationship advice. Participants rated advice satisfaction, model reliability, and helpfulness, and completed pre- and post-measures of their general attitudes toward LLMs. Overall, the results showed participants' high satisfaction with LLM-generated advice. Greater satisfaction was, in turn, strongly and positively associated with their perceptions of the models' reliability and helpfulness. Importantly, participants' attitudes toward LLMs improved significantly after exposure to the advice, suggesting that supportive and contextually relevant advice can enhance users' trust and openness toward these AI systems.

## 1 Introduction

The proliferation of Large Language Models (LLMs) has begun to reshape human interactions, extending into personal domains such as romantic relationships. While traditional relationship science has long examined how people seek support, interpret advice, and form impressions of social partners, LLMs now act as novel social agents capable of simulating these processes through natural language. Recent studies show that 26 % of the U.S. adults - including 49% of GenZ have turned to AI for dating help (Psychology Today, 2025). Recently, Chun and colleagues (2025) showed that LLMs can be structured to help participants practice conflict resolution strategies within their romantic relationships. This trend represents a significant shift in how individuals navigate intimate relationships, moving from traditional advice sources - friends, family, and professional counselors to AI-mediated guidance (Machia and Ogolsky, 2021).

From a computational linguistics perspective, relationship advice presents unique challenges for LLMs. Unlike factual question-answering or text summarization, romantic guidance requires nuanced understanding of emotions, cultural contexts, and interpersonal dynamics. The advice must balance empathy with actionability while navigating sensitive topics that can significantly impact users' well-being. Recent work by Chun and colleagues (Chun et al., 2025) showed that LLMs can be structured to help individuals practice conflict-resolution strategies with their partners, highlighting their potential as interactive tools for relationship skill development.

The adoption of AI systems in personal contexts is heavily influenced by user perceptions of credibility, usefulness, and trustworthiness (Araujo et al., 2020). In conversational AI specifically, user acceptance depends on factors including perceived competence, benevolence, and social presence (Brandtzaeg and Følstad, 2018). Moreover, gender differences (Utz, 2024) emerge consistently in human-AI interaction research, with variations in trust, perceived usefulness, and communication preferences (Hong et al., 2020). These findings suggest that user demographics may significantly moderate responses to AI-generated relationship advice. Despite growing real-world usage, empirical research on user perceptions of LLM-generated relationship advice remains limited. Existing studies focus primarily on technical performance metrics or general conversational quality, leaving fundamental questions about user experience and acceptance unanswered. This gap is particularly problematic given the sensitive nature of relationship advice and its potential psychological impact on users. To address this research gap, we present the first systematic empirical investigation of user perceptions of LLM-generated romantic relationship advice. Our study examines user satisfaction, perceived reliability and helpfulness, and attitude change following exposure to advice from two state-of-the-art models (Gemini 2.5 Pro and GPT 5.0). We also investigate gender differences in these perceptions, given their documented importance in human-AI interaction.

Specifically, we address four research questions:
**RQ1**: How satisfied are users with LLM-generated romantic relationship advice?
**RQ2**: How reliable and helpful do users perceive LLM-generated romantic-relationship advice?
**RQ3**: How do user perceptions of satisfaction, reliability, and helpfulness vary by gender?
**RQ4**: Do users' general attitudes toward LLMs change after exposure to LLM-generated advice, and do these changes vary by gender?

Our findings contribute to the growing body of research on human-AI interaction in sensitive domains and provide empirical grounding for understanding user acceptance of AI-generated personal guidance. These insights have implications for both the development of more effective conversational AI systems and the ethical deployment of LLMs in personal advisory contexts.

This paper is structured as follows: In section two, we present the details of the data collection and methodology. Next in section three, we present our analyses and key findings. We then discuss the results and their implications in section four, followed by limitations of this study (section five) and concluding remarks (section six). Finally, we provide directions for future work in section seven.

# 2 Methods

## 2.1 Participants

We collected data from 102 U.S.-based participants (51 males, 48 females, 3 non-binary; age range: 21–60 years), all fluent in English, recruited via Prolific (Prolific, 2025). All participants were provided information consent prior to beginning the study and were compensated at a rate consistent with the platform's fair wage policy. The sample was ethnically diverse: 62.75% identified as White ($n$ = 64), 16.67% as Black or African American ($n$ = 17), 8.84% as Hispanic or Latino ($n$ = 9), 4.90% as Asian ($n$ = 5), and 6.84% as Multiracial or another ethnicity ($n$ = 7). Regarding relationship status, 46.08% of participants were single ($n$ = 47), 26.47% were married ($n$ = 27), 17.65% were in a committed relationship but not married ($n$ = 18), and the remaining 9.80% were divorced, widowed, or in another arrangement ($n$ = 10). Based on this data, we now detail the methodology followed in our study.

## 2.2 Task and Procedure

We constructed a hypothetical romantic relationship scenario adapted from a Reddit post and used it to prompt two large language models (Gemini 2.5 Pro and GPT-5.0). Each model received the same prompt, scenario, and response template to standardize output format (see Appendix for details). These models were selected because they represent two state-of-the-art, publicly accessible systems known for their strong reasoning capabilities and high-quality conversational performance. The response/advice generated by each model was then presented to the participants in a 15–20 minute online survey as explained below:

## 2.3 Pre- and Post- LLM Exposure Measures

*Pre-exposure Measures:* To measure general attitudes toward LLMs, participants first completed an adapted version of the Computer Attitude Scale (Nickell and Pinto, 1986), a well-established psychometric instrument originally developed to assess attitudes toward computer technology. This scale has been widely used in human-computer interaction and social psychology research, and provides a validated framework for measuring technology acceptance and comfort. This constituted their pre-attitude responses.

*LLM evaluation:* The participants were then shown the prompt, the hypothetical romantic relationship scenario, and the corresponding response/advice generated by both models. After reviewing each LLM's response/advice, participants evaluated them by reporting their satisfaction (the extent to which the advice met their expectations and felt supportive; on a 1 = Extremely dissatisfied to 5 = Extremely satisfied scale), their self-rating of the LLM's advice (0–10 scale), and their overall assessment of perceived reliability (trustworthiness) and helpfulness (usefulness) of each LLM's model (Lankton and McKnight, 2011). Participants were not allowed to return to earlier questions to modify their responses.

*Post-exposure measures:* Finally, participants rated their general attitudes again toward LLMs' advice to assess any changes (Nickell and Pinto, 1986), provided demographics, and were debriefed (see Appendix for more details).

## 3 Data Analysis and Major Findings

Data was analyzed using paired sample t-tests and Pearson correlations to address our research questions:

*High Satisfaction with LLMs' Advice (RQ1):* Participants reported high satisfaction with advice generated by both LLMs. Although Gemini 2.5 Pro received slightly higher satisfaction scores than GPT 5.0 (See Table 1), a paired-samples t-test revealed no statistically significant difference between the two. This suggests that users were broadly satisfied with both models' advice/response (Figure 1).

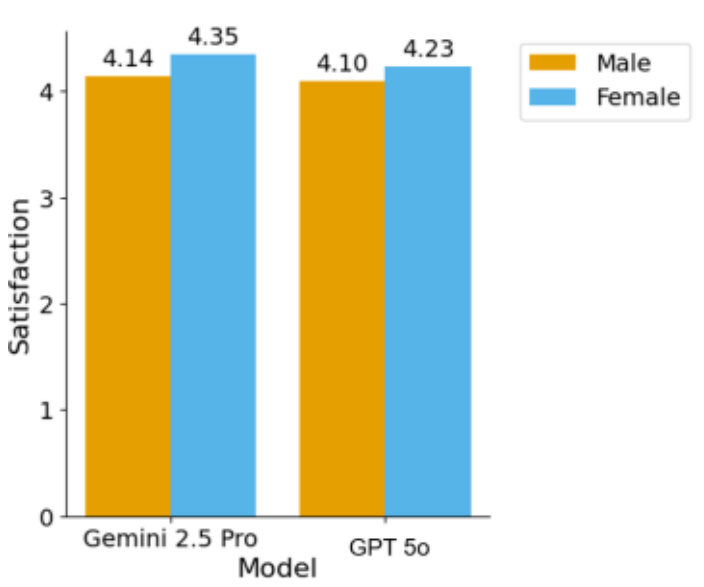

Figure 1: Participants' Average Satisfaction with LLM's Response

*LLMs' reliability and helpfulness were positively perceived by the users (RQ2):* Participants perceived the reliability and helpfulness of both LLMs favorably. Although Gemini 2.5 Pro was rated slightly higher than GPT 5.0 on these dimensions, the differences were modest (See Table 1), indicating that users viewed both models as dependable and useful sources of relationship advice.

*Gemini 2.5 Pro received slightly higher satisfaction, reliability, and helpfulness ratings across genders (RQ3)*: Both male and female participants reported high satisfaction with responses from Gemini 2.5 Pro and GPT 5.0, with Gemini 2.5 Pro receiving slightly higher satisfaction ratings overall. Across genders, both models were perceived as reliable and helpful; however, Gemini 2.5 Pro was consistently rated somewhat higher on both reliability and helpfulness compared to GPT 5.0 (See Table 2, Figures 2, 3).

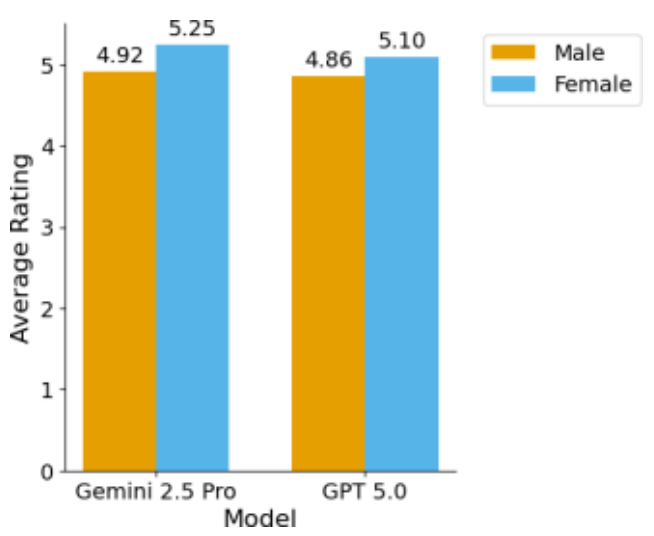

Figure 2: Reliability ratings for Gemini 2.5 Pro and GPT-5.0 responses

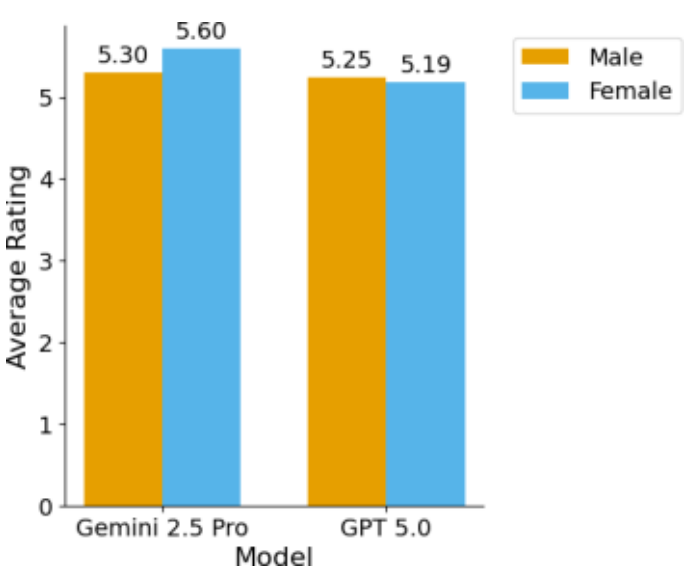

Figure 3: Helpfulness scores for Gemini 2.5 Pro and GPT-5.0 responses

RQ4:*General Attitudes became more favorably after exposure to LLM-generated advice*: We tested whether participants' general attitudes toward LLMs improved after exposure to model-generated advice.

- *Gender-Based Analysis:* An exploratory analysis by user gender showed minor positive shifts in mean attitudes for both males (pre: 4.91, post: 5.00) and females (pre: 4.98, post: 5.05) (Figure4).

- *Relationship with Satisfaction:* We found that attitude change was mediated by user satisfaction. Pre-test attitudes were a significant predictor of satisfaction with both Gemini 2.5 Pro ($r = .35$, $p < .05$) and GPT 5.0 ($r = .35$, $p < .05$). Subsequently, satisfaction was a significant predictor of post-test attitudes for both Gemini 2.5 Pro ($r = .36$, $p < .05$) and GPT 5.0 ($r = .37$, $p < .05$).

| Metric (Scale) | Gemini 2.5 Pro | GPT 5.0 |
|---|---|---|
| Satisfaction (1–5) | $M = 4.24$, $SD = 0.78$ | $M = 4.15$, $SD = 0.81$ |
| Reliability (1–7) | $M = 5.02$, $SD = 1.12$ | $M = 4.91$, $SD = 1.19$ |
| Helpfulness (1–7) | $M = 5.38$, $SD = 1.05$ | $M = 5.17$, $SD = 1.15$ |

Table 1: Users' evaluation scores for LLM advice ($N = 102$)

| Gender | Metric | Gemini 2.5 Pro | GPT 5.0 |
|---|---|---|---|
| Male | Satisfaction (1–5) | 4.13 ± 0.72 | 4.09 ± 0.64 |
| | Reliability (1–7) | 4.91 ± 1.30 | 4.86 ± 1.30 |
| | Helpfulness (1–7) | 5.30 ± 1.19 | 5.25 ± 1.20 |
| Female | Satisfaction (1–5) | 4.35 ± 0.72 | 4.22 ± 0.83 |
| | Reliability (1–7) | 5.25 ± 1.20 | 5.09 ± 1.26 |
| | Helpfulness (1–7) | 5.59 ± 1.19 | 5.19 ± 1.25 |

Table 2: Evaluation criteria scores across gender ($N = 102$). Values are presented as mean ± SD.

- *Overall Attitude Change:* A paired-samples t-test revealed a statistically significant increase in favorable attitudes from pre-test ($M = 3.68$, $SD = 0.55$) to post-test ($M = 3.91$, $SD = 0.61$). This change was significant ($t(101) = -2.73$, $p < .05$) with a medium effect size (Cohen's $d = -0.46$).

Taken together, these findings suggest that individuals who initially have more favorable views of LLMs tend to evaluate LLM-generated advice more positively, and those positive evaluations strengthen their overall attitudes even more. This pattern is consistent with a confirmation bias effect, highlighting how pre-existing dispositions shape perceptions while also being shaped by new, supportive interactions, a concept well studied in social psychology (Nickerson, 1998).

## 4 Discussion

Our preliminary study yields several key insights in the field of Human-centered AI: *Personal Context Gateway for AI Acceptance:* First, our findings show that LLMs are not only perceived as helpful and reliable sources of romantic advice but that engaging with them in a personal, relatable context can shift broader attitudes toward these AI systems. While public discourse on AI is often dominated by abstract fears or complex technical capabilities, our results indicate that a single, helpful interaction on a familiar human problem can be a powerful driver of user trust and acceptance of LLMs in emotionally sensitive domains.

*Gender Differences in User Perception:* Although satisfaction with LLM responses was high across all genders, there were small but consistent differences in perceived reliability and helpfulness. Female participants rated Gemini 2.5 Pro slightly higher than GPT 5.0 on both reliability and helpfulness, whereas male participants showed a smaller distinction between the models. These findings suggest that gender can subtly influence evaluations of AI advice, highlighting the importance of considering demographic factors when assessing human–AI interactions in intimate or emotionally salient contexts.

*Reinforcing Loop of Trust and Bias:* The correlation between prior beliefs and user experience is a known phenomenon in Human-Computer Interaction (HCI), but its operation in an intimate domain is particularly salient. We found that pre-existing attitudes toward LLMs were positively associated with satisfaction ratings of LLM advice ($r = .35$ for both models), and that satisfaction ratings were, in turn, positively associated with post-test attitudes ($r \approx .37$ for both models). Together, these findings point to a reinforcing cycle: users who begin with favorable views of LLMs become more trusting after positive interactions, while skeptical users may resist or disengage if their first interaction was negative. This highlights the "first mile" of user experience as a critical determinant of long-term acceptance of LLMs in sensitive domains such as relationship

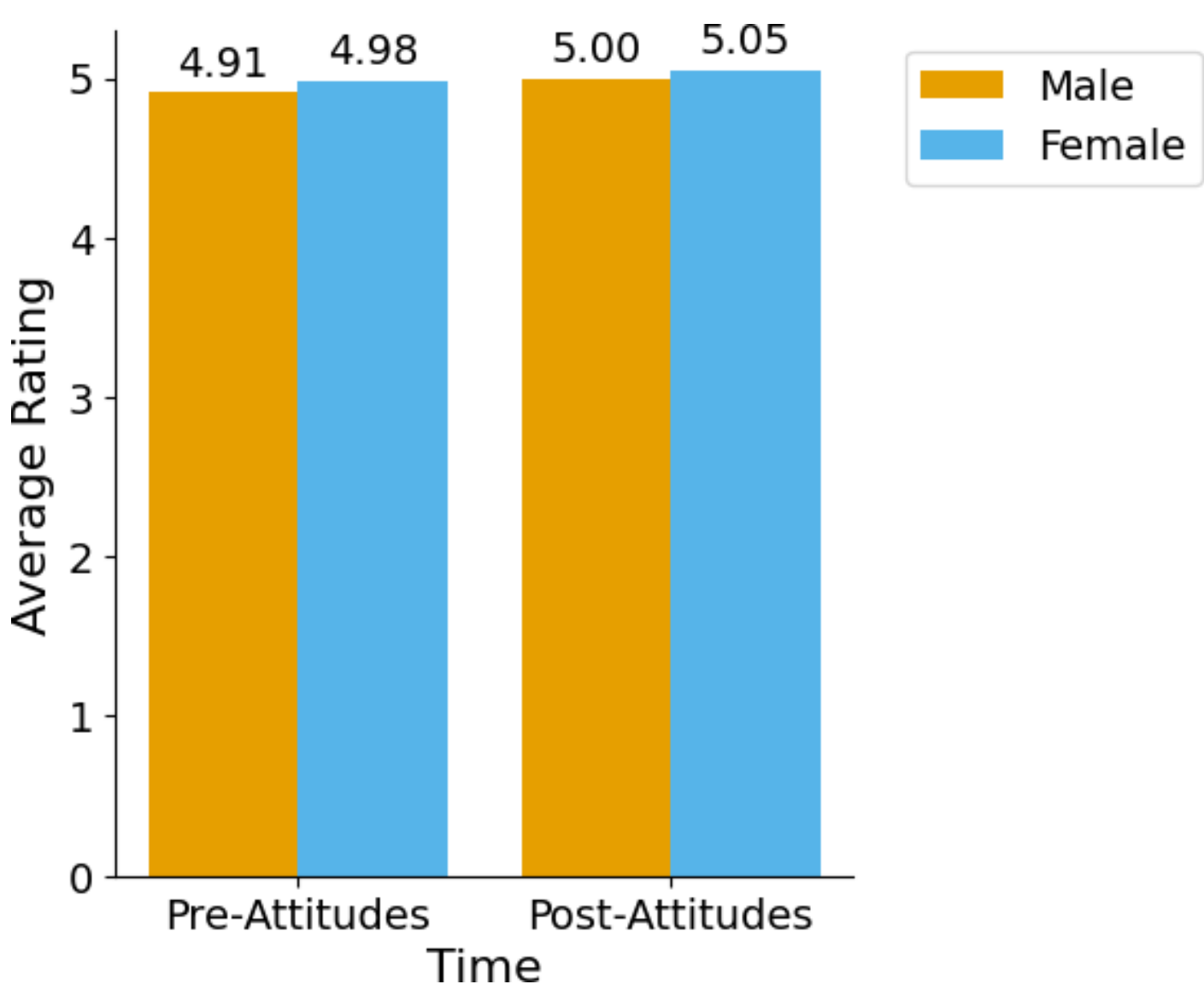

Figure 4: Participants' Average Attitude Towards LLM's Response

support.

*Direct Challenges for AI design and ethics:* The relative ease with which users' attitudes can be shifted underscores developers' responsibility. LLMs providing relationship advice should be designed with *epistemic humility* - acknowledging limitations, avoiding overconfidence, and pointing users to qualified human support when appropriate. Embedding such safeguards is essential to ensure trust is built responsibly in high-stakes interpersonal domains.

## 5 Limitations

Several limitations should be considered while interpreting these findings. First, this study used only two LLMs - Gemini 2.5 Pro and GPT-5.0, which limits the generalizability of results. Second, this study captures only immediate attitude shifts following brief exposure to LLM advice, which may not reflect how attitudes evolve with sustained use over time. Third, the use of a single hypothetical romantic relationship may not capture the emotional complexity and personal investment inherent in real relationship challenges, potentially limiting the ecological validity of our findings. Fourth, our adaptation of the Nickell-Pinto Computer Attitude Scale (1986), while providing a validated framework, was originally designed for general computer technology and may not fully capture the nuances of attitudes toward modern conversational AI systems. Fifth, prior work shows that individuals often express greater trust in algorithmic recommendations when they apply advice to external targets or generalized situations, but become more skeptical and sometimes resistant when the same advice is personally self-relevant (Morewedge, 2022). Thus, while our participants evaluated the LLM-generated advice positively, the judgments may not fully generalize to scenarios where individuals receive advice that is explicitly personal or directly tied to their own relationship decision-making. Additionally, the sample consisted primarily of U.S. based participants with specific demographic characteristics and access to technology, which may limit generalizability across cultures and populations. The absence of human-generated advice as a comparison baseline also limits our ability to contextualize LLM performance relative to traditional advice sources. Finally, the controlled survey environment differs substantially from authentic advice-seeking contexts where individuals face genuine emotional stakes and relationship consequences.

## 6 Conclusion

This study presents the first systematic empirical investigation of user perceptions of LLM-generated romantic relationship advice, addressing a critical gap in our understanding of human-AI interaction in sensitive personal domains. Through a controlled evaluation with 102 participants examining responses from Gemini 2.5 Pro and GPT 5.0, we provide novel insights into user acceptance and attitude formation toward AI-generated guidance. Our findings reveal that users express high satisfaction with LLM-generated relationship advice, with no significant differences between the two state-of-the-art models examined. Users perceive AI-generated advice as both reliable and helpful, suggesting that current LLMs have achieved sufficient quality to earn user trust in this sensitive domain. Notably, we observed significant gender differences in perceptions, with implications for the design and deployment of conversational AI systems in personal advisory contexts. The study also demonstrates that exposure to LLM-generated advice can produce measurable changes in users' general attitudes toward AI systems, indicating that domain-specific interactions shape broader AI perceptions. This finding has important implications for understanding how user experiences with LLMs in personal contexts influence their overall acceptance of AI technologies. From a computational linguistics perspective, these results suggest that modern LLMs have developed sufficient capability in generating contextually appropriate, empathetic responses for relationship guidance—a complex NLP task requiring nuanced understanding of human emotions and social dynamics. The comparable performance between models indicates that the underlying transformer architectures and training approaches have converged on effective strategies for this challenging domain. While these findings advance our understanding of human-AI interaction in personal domains, the limitations identified above highlight important avenues for future research to enhance both the validity and practical applicability of these insights.

## 7 Future Work

Future research should systematically address the limitations identified in this study through several key avenues. Incorporating additional LLM models and fine-tuned variants would allow for broader generalization across models and training paradigms. To address temporal concerns, longitudinal research examining the stability and evolution of user attitudes across repeated, real-world interactions with LLMs would offer deeper insight into the durability and authenticity of the observed effects. Moving toward greater ecological validity, researchers should investigate real-world advice-seeking interactions with appropriate ethical safeguards, including partnerships with relationship counselors and robust consent procedures. Expanding generalizability, cross-cultural research examining how cultural values and relationship norms influence perceptions of AI-generated advice would enhance external validity. Studies with more diverse populations, including different age groups (e.g., teenagers) and technological backgrounds, would further broaden applicability. Improving measurement and comparison, future work should develop and validate attitude scales specifically designed for modern conversational AI systems rather than adapting decades-old computer attitude measures. Comparative studies including human-generated advice baselines, different LLM models, and various relationship scenarios would provide essential context for interpreting AI advice effectiveness. Deepening understanding of mechanisms, researchers should identify specific qualities of AI-generated advice (e.g., empathy, actionability, personalization) that drive user satisfaction and attitude change through more granular content analysis. This would inform the development of more effective AI advisory systems. Investigating real-world impact, perhaps most critically, future studies should examine the behavioral outcomes of AI advice implementation—whether users actually follow AI recommendations and with what results. This represents a crucial step for understanding the practical impact of LLM-generated relationship guidance beyond user perceptions. Together, these research directions would build upon our initial findings to create a more comprehensive understanding of AI's role in personal advisory contexts and its broader implications for human-AI interaction in sensitive domains.

# A Data Collection and Dataset

We recruited a total of 102 participants for this study through the online research platform Prolific. Inclusion criteria required participants to be based in the United States, be at least 18 years of age, and be fluent in English. All participants provided informed consent before beginning the study and were compensated for their time at a rate consistent with the platform's fair wage policy. The final sample consisted of 51 males, 48 females, and 3 non-binary individuals. The age of participants ranged from 21 to 60 years, with a mean age of 39.66 (M=39.66,SD=10.43). The sample was ethnically diverse: 62.75% of participants identified as White (n=64), 16.67% as Black or African American (n=17), 8.84% as Hispanic or Latino (n=9), 4.90% as Asian (n=5), and 6.84% identified as Multiracial or another ethnicity (n=7). Regarding relationship status, 46.08% of participants were single (n=47), 26.47% were married (n=27), 17.65% were in a committed relationship but not married (n=18), and the remaining 9.80% were divorced, widowed, or in another arrangement (n=10).

## A.1 Ethical Statement

All participants provided informed consent and were fully briefed on the study's procedure. The participation was voluntary and a fair compensation was provided for this study.

## A.2 Prompt provided to the LLM

**Prompt:**

You are a trusted support resource giving relationship advice to a university student (18–25 yrs old) who shared their story with you. You are supposed to offer 2–3 concrete, actionable advice to address their issue. Be supportive, empathetic, and practical. Do not diagnose. Be non-judgmental. Keep your advice under 100 words. Also, provide your reasoning behind your advice.

**Story:** One of our shared interests was video games. So, after that first week, we started playing games together and talking on Discord. We would talk on Discord every day whenever we were both free and stay up super late just learning about each other and telling each other stories from our childhood. She's from Texas, and I'm from California, so our time difference is two hours. She told me her schedule for that semester of college, and I would make sure to wake up early enough to message her at least for a little bit before class.

Fast forward about a month, and we're both pretty flirty with each other when we're talking. I'm not sure if we were doing it on purpose or if we were even aware of it at the time. I had learned that she had a boyfriend, but she was super unhappy with their relationship, which was why she needed to talk to somebody in the first place. I also learned that he had cheated on her very early on and that she took him back because she was scared he'd kill himself if she didn't.

So one night, while we were done playing games and just hanging out and talking on Discord, I just asked her point-blank if she thought her relationship with her ex was going anywhere. She said that she didn't think so, so I asked if she thought that the two of us had a chance.

We've been together for just over a year now. She's graduated from college and has a job. She has to be up at 6:30 CST every morning for work so I set an alarm for 4:30 AM PST so I can be awake, at least for a few minutes, to talk to her and tell her to have a good day. I've been over there to visit her every month since July last year and once she's been at her job for six months, she can start taking vacation time and visit me.

I've been thinking about taking a big step—moving in together. But there's this lingering thought: she hasn't visited me yet, and sometimes she seems unsure about where she wants this relationship to go. It's surreal to think about how we met in such an unconventional way and how deeply I've grown to love her. I cherish every moment we share, and I want to take things to the next level. But now I'm anxious—what if she doesn't feel the same? What if this isn't going where I hoped it would?

**Exact Output Format:** My advice: <action 1><action 2>

My reasoning (less than 100 words; keep it brief)

Your self-rating on how assured you think the user would feel? [return only one numerical between 1 (no confidence) and 7 (very confident)]

## A.3 Outputs given by LLMs

Below are the outputs provided by Gemini 2.5 Pro (Comanici et al., 2025) and GPT 5.0 (OpenAI, 2025) that were provided to the participants:

**Gemini 2.5 Pro:**
My advice: Have an open and honest conversation. Share your feelings and hopes for moving in together, but also create a safe space to ask her, without pressure, where she sees the relationship going. Focus on planning her first visit to you as the next concrete step. Experiencing your life, home, and environment is a crucial milestone for both of you before considering cohabitation.

My reasoning (less than 100 words; keep it brief) Direct communication is the best way to address your anxiety and understand her perspective. Her visiting you is a practical and necessary step in closing the distance. This allows you both to see how well your lives fit together in your space before you make a significant commitment like moving. This approach validates your feelings while respecting the relationship's current stage.

Your self-rating on how assured you think the user would feel? [return only one numerical between 1(no confidence) and 7 (very confident)]: 6

**GPT 5:**
My advice: Have an open conversation with her about your hopes for moving in together and ask her directly about her feelings toward the future of the relationship.

Suggest that she visit you before making any big decisions—seeing your life in California will give both of you a clearer perspective.

Take things one step at a time—set small goals (like planning her first visit) before committing to a major move.

My reasoning: Clear communication and gradual steps help reduce anxiety and ensure both partners feel ready and aligned before a big life change.
Your self-rating on how assured you think the user would feel? 6

## A.4 Scales and Measures

**LLM's Response Satisfaction:**
To assess satisfaction levels towards the LLM's romantic relationship advice, participants rated their satisfaction with each model's response to the story using a single item: "How satisfied are you with [Gemini 2.5 Pro's/GPT 5.0's] response to the story?" The responses were given on a 5-point scale (1 = Extremely Dissatisfied, 5 = Extremely Satisfied).

**LLM's Response Rating:**
Participants answered 1 item, "How much would you rate [Gemini 2.5 Pro's or GPT 5.0's] response to the story (out of 10)?"

**LLM's Reliability and Helpfulness:**
Following exposure to the advice, reliability, and helpfulness were measured using items adapted from Lankton and McKnight (2011). Participants answered/rated five reliability items ($\alpha$ = .34) and five helpfulness items ($\alpha$ = .07) on a 7-point scale (1 = Strongly Disagree, 7 = Strongly Agree).

**Attitudes Towards LLMs:**
To capture the attitude perceptions, participants rated 20 items about the LLM's response on a 7-point scale (1 = Strongly Disagree, 7 = Strongly Agree). The statements which captured both positive and negative attitudes toward LLMs were adapted from the Computer Attitude Scale [CAS (Nickell and Pinto, 1986)] and demonstrated a strong internal consistency ($\alpha$ = .89).

## A.5 Correlations between variables in the study

| Variable 1 | Variable 2 | r | 95% CI | p |
|---|---|---|---|---|
| Avg_Att_Pre | SatGem | .35 | [.16, .51] | .010* |
| Avg_Att_Pre | SatGpt | .35 | [.16, .51] | .011* |
| Avg_Att_Pre | RGPT_num | .31 | [.12, .48] | .045* |
| Avg_Att_Pre | Avg_reliable_Gem | .44 | [.27, .59] | <.001*** |
| Avg_Att_Pre | Avg_helpful_Gem | .46 | [.29, .60] | <.001*** |
| Avg_Att_Pre | Avg_helpful_GPT | .45 | [.29, .60] | <.001*** |
| Avg_Att_Pre | Avg_reliable_Gpt | .49 | [.32, .62] | <.001*** |
| Avg_Att_Pre | Avg_Att_Post | .96 | [.93, .97] | <.001*** |
| SatGem | RGem_num | .75 | [.65, .83] | <.001*** |
| SatGem | Avg_reliable_Gem | .54 | [.38, .66] | <.001*** |
| SatGem | Avg_helpful_Gem | .72 | [.61, .80] | <.001*** |
| SatGem | Avg_Att_Post | .36 | [.18, .52] | .006** |
| RGem_num | Avg_reliable_Gem | .56 | [.41, .68] | <.001*** |
| RGem_num | Avg_helpful_Gem | .64 | [.50, .74] | <.001*** |
| RGem_num | Avg_helpful_GPT | .32 | [.13, .49] | .042* |
| RGem_num | Avg_reliable_Gpt | .31 | [.12, .48] | .046* |
| SatGpt | RGPT_num | .81 | [.73, .87] | <.001*** |
| SatGpt | Avg_helpful_Gem | .39 | [.21, .54] | .002** |
| SatGpt | Avg_helpful_GPT | .66 | [.53, .75] | <.001*** |
| SatGpt | Avg_reliable_Gpt | .57 | [.42, .68] | <.001*** |
| SatGpt | Avg_Att_Post | .37 | [.19, .53] | .004** |
| RGPT_num | Avg_helpful_GPT | .65 | [.52, .75] | <.001*** |
| RGPT_num | Avg_reliable_Gpt | .61 | [.46, .72] | <.001*** |
| RGPT_num | Avg_Att_Post | .36 | [.17, .52] | .009** |
| Avg_reliable_Gem | Avg_Att_Post | .47 | [.31, .61] | <.001*** |
| Avg_helpful_Gem | Avg_helpful_GPT | .68 | [.56, .78] | <.001*** |
| Avg_helpful_Gem | Avg_reliable_Gpt | .63 | [.49, .73] | <.001*** |
| Avg_helpful_Gem | Avg_Att_Post | .50 | [.34, .63] | <.001*** |
| Avg_helpful_GPT | Avg_reliable_Gpt | .85 | [.79, .90] | <.001*** |
| Avg_helpful_GPT | Avg_Att_Post | .53 | [.37, .65] | <.001*** |
| Avg_reliable_Gpt | Avg_Att_Post | .55 | [.40, .67] | <.001*** |

Table 3: Correlations (r) between variables in this study (Att: Attitude; Pre: Pre-test rating; Post - Post-test rating; Sat: Satisfaction; reliable: Reliability; helpful- Helpfulness; Gem: Gemini 2.5 Prom; GPT/Gpt: GPT 5.0; RGPT: GPT 5.0 Ratings)